# Government Initiatives: The Missing Key for E-commerce Growth in KSA

R. AlGhamdi, S. Drew and S. Alkhalaf

*Abstract*—This paper explores the issues that influence online retailing in Saudi Arabia. Retailers in Saudi Arabia have been reserved in their adoption of electronically delivered aspects of their business. Despite the fact that Saudi Arabia has the largest and fastest growth of ICT marketplaces in the Arab region, e-commerce activities are not progressing at the same speed. Only very few Saudi companies, mostly medium and large companies from the manufacturing sector, are involved in e-commerce implementation. Based on qualitative data collected by conducting interviews with 16 retailers and 16 potential customers in Saudi Arabia, several factors influencing online retailing diffusion in Saudi Arabia are identified. However, government support comes the highest and most influencing factor for online retailing growth as identified by both parties; retailers and potential customers in Saudi Arabia.

*Keywords*—government support, key factor, online retailing growth, Saudi Arabia

## I. INTRODUCTION

MANY businesses around the world have introduced e-commerce models as part of their operations, seeking the many advantages that the online marketplace can provide [1]. Since 2000, e-commerce's rapid growth is obvious in the developed world. Global e-commerce spending has currently reached US$10 trillion and was US$0.27 trillion in 2000 [24]. Despite the fact that Saudi Arabia has the largest and fastest growth of ICT marketplaces in the Arab region [4], [5], [6] and [7], e-commerce activities are not progressing at the same speed [5], [8], [9], [10], and [11]. Comparing electronic commerce in Saudi Arabia to the global volume, it is low at $150 million [6]. By the end of 2007, the number of officially registered Saudi commercial organizations is 695,200; however, Saudi Arabia is still a late adopter in e-commerce field [11]. Only 9% of Saudi commercial organizations, mostly medium and large companies from the manufacturing sector, are "involved in e-commerce implementation [10]. The question which should be asked why Saudi Arabia is a late a adopter of online retailing systems and how the adoption can be accelerated?

R. AlGhamdi is with school of Information and communication Technology, Griffith University, 170 Kessels Rd, Nathan QLD 4111 (e-mail: rayed.alghamdi@griffithuni.edu.au).
S. Drew is with school of Information and communication Technology, Griffith University, 170 Kessels Rd, Nathan QLD 4111 (e-mail: s.drew@griffith.edu.au).
S. Alkhalaf is with school of Information and communication Technology, Griffith University, 170 Kessels Rd, Nathan QLD 4111 (e-mail: s.alkhalaf@griffith.edu.au)

## II. REVIEW OF E-COMMERCE IN KSA

Since the responsibility of e-commerce has transferred to the Saudi Ministry of Communications and Information Technology in 2005/06, official/government information about e-commerce in Saudi Arabia is poor. Before that date the Saudi Ministry of Commerce established a permanent technical committee for e-commerce including members from the Ministries of Commerce, Communication and Information Technology and Finance. It also includes members from the Saudi Arabian Monetary Authority and King Abdulaziz City for Science and Technology [4]. The roles of this committee are to follow the developments in the field of e-commerce and take the necessary steps to keep pace with them. The committee will learn from international experiences in this area, identifying the particular needs and requirements to take advantage of e-commerce, following-up to completion of the development work required, and the preparation of periodic reports on progress of work on a regular basis [4]. The committee has prepared a general framework for a plan to apply e-commerce systems in Saudi Arabia. This framework includes the improvement of various factors involved with e-commerce transactions (e.g. IT infrastructure, payment systems, security needs, legislations and regulations, delivery systems etc). The plan also includes the development of e-commerce education and training [4]. This information gained from the only source in this regard published by Ministry of commerce. However no further information about can be gained about specific projects for e-commerce development in the country.

While United States, followed by Europe, constitutes the largest share with about 79% of the global e-commerce revenue [24]. However, the African and Middle East regions have the smallest share with about 3% of the global e-commerce revenue [24]. For Saudi Arabia, several studies have been conducted to investigate the challenges of e-commerce in Saudi Arabia. These challenges involve the absence of clear e-commerce regulations, legislation, and rules [13] and [8]. Although Saudi Arabia contributes to the efforts of UNCITRAL (United Nations Commission into International Trade Laws) [4], there is a need to have major development in terms of e-commerce regulations, legislations and rules to protect the rights of all parties involved in e-commerce transactions [8], [13] and [11].

Reference [9] conducted a study to report "certain Internet characteristics and e-commerce issues" and come up with challenges that face "the diffusion of the Internet and its application" in the Arab region. Regarding applying e-commerce, he found that there are primary concerns for both business owners/mangers and online customers. The Arab business managers take into their consideration "technical obstacles and the attitudes and behaviours of e-commerce consumers". The Internet customers mentioned business's reputation, security, privacy, and legal regulations.

Some studies have been conducted to figure out what makes Saudi Arabia a late adopter in the field of e-commerce. A commercial guide for US companies 'doing business in Saudi Arabia' [6] stated some of the barriers that were encountered in the quick implementation of electronic business in the Saudi markets. These inhibitors involved the slowness of Internet services, resistance to social adaptation to a new commercial paradigm, lack of trust of online business, Shortage of skilled employees for implementation and maintenance of e-business systems [6] and [12].

In 2007, Saudi Communication and Information Technology Commission (CITC) carried out an extensive study to evaluate the current situation of the Internet, and various aspects involving the Internet usage in Saudi Arabia. One of these aspects is e-commerce awareness and activities. As introduced earlier, for the business commercial organizations, they have found that only 9% of Saudi commercial organizations, mostly medium and large companies from the manufacturing sector, are "involved in e-commerce implementation". It is reported that only 4 out of 10 private companies have their own website. This percentage takes on a higher proportion for the larger oil, gas and manufacturing companies. For the users (customer), they have found that 43% of the respondents were aware of e-commerce and only 6% ever bought or sold products online, "mainly airline tickets and hotel bookings" [10].

Reference [7] said: "Universally, e-commerce has three main 'pillars': communication and Internet, payment, and delivery services". Looking at delivery services in Saudi Arabia, the Saudi postal services remain inefficient (Alfuraih 2008). From the establishment of Saudi Post until 2005, individuals had no uniquely identifying home addresses and the mail was not delivered to homes and offices (Saudi Post 2008). If individuals want to receive mail, they have to subscribe to have mail boxes in the local post offices [7]. In 2005, the new project for addressing and delivery to homes and buildings was announced and approved by Saudi Post [7] and [14].

III. RESEARCH METHODOLOGY

As this study initially involves exploratory research, a qualitative approach is adopted. Qualitative approach is suitable to discover and gain in-depth understanding of the phenomena [16]. A semi-structure interview was designed based on the Diffusion of Innovation theory (DOI) [17]. Reference [17] identified five attributes determining the innovation's rate of adoption. He highly recommended that each diffusion research should develop the measures of the five perceived attributes. These five variables are (1) perceived attributes of innovations (Relative advantage, Compatibility, Complexity, Trialability, and Observability), (2) type of innovation-decision (optional, collective, authority), (3) communication channel diffusing the innovation at various states in the innovation-decision process (mass media, interpersonal), (4) nature of the social system (norms, degree of network interconnectedness, etc), and (5) extent of change agent's promotion efforts [17].

Interviews were conducted with 16 retailers' decision makers (owner, headquarter manager, marketing manager, and IT manager) and 16 Saudi citizens (8 males and 8 females). The sample of retailers decision makers were selected to cover large, medium and small companies and also to cover different types of retail businesses (telecommunications, computers, sports, super markets, restaurants, printing services, Internet services, electrical and electronic products, beauty and body cares, books, watches and clocks, and Chocolate and Biscuit manufacturing). Interview questions, answers and discussions for both retailers and customers were all in Arabic, except with retailers were in English, and the researcher translated the transcripts into English.

IV. RESULTS AND DISCUSSION

The study come up with several issues negatively and positively affects the online retailing growth in Saudi Arabia. The sample from retailers and potential customers came up with common issues and some other issues are concerning a party but not another.

The common problems that both parties identified include habit of people in selling and buying in KSA; issues involved with delivery; issues involved with online payments; lack of strong ICT infrastructure; no clear regulations and legislations; Lack of trust; and IT knowledge. In addition to the previous list of concerns, there are several issues concern only one party and not another. The issues that can be identified only from retailers' perspectives that impeded online retailing growth include setting up cost; resistance to change; difficulty to offer competitive advantage; and type of products is prevent to sell online. The issues that only concerns customers include absence of owning houses' mailboxes and absence of online retailers for most need products.

The common positive factors that shared between the sample of retailers and customers include government support; offering trustworthy and secure online payment options; educational programs; enhance ICT infrastructure; provision of sample e-commerce software to trial. In addition to these factors, customers see other factors that support them to buy online including competitive prices; e-retail's website proficiency; owning houses addresses; and the advantage of bricks-and-clicks model. And then the study comes up with seven key drivers to online retailing growth in KSA [22].

However, most of the participants linked whether the negative or positive issues by somehow with the government support. It is not surprising for Saudis to seek government's help. This is because Saudi government plays a central role and people have a tendency to trust what comes through the government. The role that the government can responsibly play in facilitating e-retail is based around three key functions: facilitation, supervision, and control.

The Saudi government role to regulate, support, facilitate and control e-commerce activities seems to be missing for the following reasons.

(1) In 2001, the Ministry of Commerce established a permanent technical committee for e-commerce including members from the Ministries of Commerce, Communication and Information Technology and Finance. It also includes members from the Saudi Arabian Monetary Authority (SAMA) and King Abdulaziz City for Science and Technology (KACST) [4]. However, very few information about this committee has been founded. A phone call made on 13 Dec 2010 to Saudi Ministry of Commerce to gain more information about this committee. A ministry representative replied "this committee does no longer exist under Ministry of Commerce. Since 2006, the responsibility of e-commerce has transferred to the Ministry of Communications and Information Technology".

(2) In 2006, Saudi Ministry of Communications and Information Technology produced a national plan for ICT in Saudi Arabia entitled "the national communications and information technology plan". The vision statement of this plan is "the transformation into an information society and digital economy so as to increase productivity and provide communications and IT services for all sectors of the society in all parts of the country and build a solid information industry that becomes a major source of income" [18]. They formulated seven general goals to realize this vision. These seven objectives further detailed into 26 specified objectives, 62 implementation policies, and 98 projects [18] and [19]. However, this plan does not include specific details involve e-commerce. The seven main goals, 26 specified objectives, 62 implementation policies, and 98 projects mainly involve with (public sector) e-government and e-learning and what need to be done to support its activities. No details provided to show the current situation of e-commerce in Saudi Arabia and what need to be done in the future to facilitate and support its activities.

(3) King Abdulaziz City for Science and Technology (KACST) and Saudi Ministry of Economy and Planning [20] released a document entitled "strategic priorities for information technology program". This document is a specific plan for the needed researches in IT to support the national communications and information technology plan. It lists key needs in Saudi Arabia to support IT growth. Among these priorities, e-commerce applications, activities, and researches are not listed with exceptional to authentication for e-government and e-commerce.

(4) In 2005, the new project for addressing and delivery to homes and buildings was announced and approved by Saudi Post[7] and [14]. 'Wasel' is a mail service enables the residents in Saudi Arabia to receive all their mail at their residence, delivered to their mail box free of charge. Residents in Saudi Arabia have to give a phone call to Saudi Post phone number, visit a Saudi post office or apply online to apply for getting a mailbox with home physical address. There is also another service called "Wasel Special". This service is chargeable includes beside to the delivering mail service other six services: sending mail from a home mailbox, delivery with e-stamp, temporary safe keeping, temporary forwarding, P.O.Box transfer and e-mail notification. However, this service is not covering all cities in Saudi Arabia, till now only main cities[20]. The number of subscribers in the 'Wasel' service reached more than half a million [21]. This number represents almost %2 of population who own individual houses mailbox. While this service still new in the Saudi environment which consumes time to be recognized and adopted by most of the population; however, Saudi Arabia comparing to developed world is too late providing individual addresses. While the service of providing individual houses mailboxes is there, however, the percentage of population owns a mailbox ring a bell for more efforts in order to investigate what inhibit citizens to own home mailbox. Problems might be related to the lack of awareness issues of the importance mailbox, or might be related to real problem that the citizens do not trust receiving their mails through this new service. Another issue may be raised here that among the participants in this research do not know that there is a service delivering mail to individual houses or do not know that there are direct addresses for their houses with numbers and streets names. In all cases, while the service is there, more efforts are needed to motivate the citizens owning house mailboxes and figure out the problems that they face.

On 23rd Dec 2010, a newspaper article stating all these facts about the government missing role was published in Alriyadh [23], the main newspaper in Saudi Arabia. Publishing this article was for two reasons: (1) to make an impact of this study contributing to the development by identifying the current problems to find solutions, and (2) to get feedback and response from the mentioned government sectors above. Fortunately, a representative from Saudi Communications and Information Technology Commission (CITC) contacted the researcher in response to this article. He told that we still in early stage. Currently we are conducting a survey in e-commerce in Saudi Arabia and a report may be published in May/ June 2011. The researcher is currently conducting interview with CITC to find out more information.

So, a preliminary model is designed based on the findings showing the central role of the government enhancing the growth of online retailing in Saudi Arabia. However, this model is in its early stage as another step of this study will be conducted. A quantitative approach (survey) will be conducted to confirm the current findings (or coming up with new) in a wider sample in a range of participants in Saudi Arabia.

## V. Conclusion and Future Work

This paper has investigated the issues that positively and negatively influence online retailing growth in Saudi Arabia. It comes up with several issues inhibiting and driving towards online retailing in Saudi Arabia. However the government support seems to be the missing key for online retailing support and growth in Saudi Arabia. Therefore, policy makers and developers should pay attentions to these factors and the Saudi government should take serious role to facilitate online retailing growth in KSA. However, this study is still in progress. We hope in due course to be able to develop a model in order to contribute to e-commerce development in Saudi Arabia.

## References

[1] Laudon, K.C. and C.G. Traver, *E-commerce: business, technology, society*. 3 ed. 2007, New Jersey: Pearson Prentice Hall.
[2] CIA (Central Inelegance Agency). *The World Fact Book: Saudi Arabia*. 2009 [cited 2010 20 Oct]; Available from: https://www.cia.gov/library/publications/the-world-factbook/geos/sa.html.
[3] Information Centre - Saudi Ministry of Commerce, *E-commerce in the Kingdom of Saudi Arabia* in *Arab Organization for Industrial Development Conference* 2006: Tunisia. p. 1-27.
[4] Saudi Ministry of Commerce, *E-commerce in the kingdom: Breakthrough for the future*. 2001, Rayadh.
[5] Al-Otaibi, M.B. and R.M. Al-Zahrani, *E-commerce Adoption in Saudi Arabia: An Evaluation of Commercial Organizations' Web Sites*. 2003, King Saud University: Riyadh.
[6] U.S. Department of Commerce, *Doing Business In Saudi Arabia: A Country Commercial Guide for U.S. Companies*. 2008, U.S. & Foreign Commercial Service and U.S. Department of State.
[7] Alfuraih, S. *E-commerce and E-commerce Fraud in Saudi Arabia: A Case Study*. in *2nd International Conference on Information Security and Assurance* 2008. Busan, Korea: Institute of Electrical and Electronics Engineers (IEEE)
[8] Albadr, B.H., *E-commerce*. Science and Technology, 2003(65): p. 14-19.
[9] Aladwani, A.M., *Key Internet characteristics and e-commerce issues in Arab countries*. Information and Management, 2003. 16(1): p. 9-20.
[10] CITC (Communications and Information Technology Commission), *Internet Usage Study in the Kingdom of Saudi Arabia* 2007, Communications and Information Technology Commission: Riyadh.
[11] Agamdi, A *e-Commerce Implementation Challenges and Success Factors in the Kingdom of Saudi Arabia*, in *19th National Computer Conference: the digital economy and ICT industry*. 2008: Riyadh.
[12] CITC (Communications and Information Technology Commission) , *Marketplace of Telecommunications and Information Technology in KSA* 2006, Communications and Information Technology Commission: Riyadh, p. 1-57.
[13] Al-Solbi, A. and P.J. Mayhew, *Measuring E-Readiness Assessment in Saudi Organisations Preliminary Results From A Survey Study*, in *From e-government to m-government*, I. Kushchu and M.H. Kuscu, Editors. 2005, Mobile Government Consortium International LLC: Brighton, UK. p. 467-475.
[14] Saudi Post. *Saudi Post: Establishment and Development*. 2008 [cited 2009 21 Nov]; Available from: http://www.sp.com.sa/Arabic/SaudiPost/aboutus/Pages/establishmentanddevelopment.aspx.
[15] Alrawi, K.W. and K.A. Sabry, *E-commerce evolution: a Gulf region review*. Int. J. Business Information Systems, 2009. **4**(5): p. 509-526.
[16] Neuman, W.L., *Social Research Methods: Qualitative and Quantitative Approaches*. 6th ed. 2006, Boston, MA: Pearson Education.
[17] Rogers, E.M., *Diffusion of Innovations*. Fifth ed. 2003, New York: Simon & Schuster. 551.
[18] Suadi Ministry of Communication and Information Technology, *The National Communications and Information Technology Plan*. 2006, Ministry of Communication and Information Technology: Riyadh. p. 91.
[19] Saudi Ministry of Communication and Information Technology, *The Annual Report for the National Communications and Information Technology Plan*. 2009, Ministry of Communication and Information Technology: Riyadh
[20] Saudi Post. *Production and Services*. 2010 [cited 2010 14 Dec]; Available from: http://www.sp.com.sa/English/SaudiPost/ProductsServices/Pages/Wasell.aspx.
[21] Alriyadh. *No E-government without the application of mailing Addressing System* 2010 [cited 2010 14 Dec]; Available from: http://www.sp.com.sa/arabic/news/pages/newsdetails.aspx?ItemID=419
[22] AlGhamdi, R & Drew, S 2011, 'Seven Key Drivers to Online Retailing in KSA', in P Kommers & P Isaías (eds), *Proceedings of the IADIS International Conference on e-Society 2011*, Avila, Spain, pp. 237-44.
[23] AlGhamdi, R., *E-commerce in the Kingdom and the missing hope!*, in *Alriyadh*. 2010: Riyadh; Available from: http://www.alriyadh.com/2010/12/23/article588081.html
[24] Kamaruzaman, KN, Handrich, YM & Sullivan, F 2010, 'e-Commerce Adoption in Malaysia: Trends, Issues and Opportunities', in R Ramasamy & S Ng (eds), *ICT Strategic Review 2010/11 E-commerce for Global Reach* PIKOM (The National ICT Association of Malaysia), Putrajaya, Malaysia, pp. 89-134.